\documentclass[review]{elsarticle}

\usepackage{textcomp}
\usepackage{lineno}
\usepackage{amssymb}
\usepackage{multirow}
\usepackage{subcaption}

\usepackage{xcolor}

\usepackage[colorlinks,citecolor=blue,urlcolor=blue,linkcolor=blue]{hyperref}

\journal{Journal of Cosmology and Astroparticle Physics}

\begin{document}

\begin{frontmatter}

\title{A pulsed compact low-background X-ray generator}


\author[CEA]{I.~Giomataris}
\author[CEA]{F.~Belloni\corref{cor}}
\ead{francesca.belloni@cea.fr}
\author[CEA]{F.J.~Iguaz\fnref{Iguaz}}
\author[CEA]{J.P.~Mols}
\author[CEA]{T.~Papaevangelou}
\author[CEA]{L.~Segui}

\cortext[cor]{Corresponding author}
\address[CEA]{IRFU, CEA, Universit\'e Paris-Saclay, F-91191 Gif-sur-Yvette, France}
\fntext[Iguaz]{Now at Synchrotron Soleil, BP 48, Saint-Aubin, 91192 Gif-sur-Yvette, France.}



\cortext[F. Belloni]{francesca.belloni@cea.fr}


\begin{abstract}
A novel type of  pulsed X-ray generator especially suitable for rare event searches, as dark matter and neutrino experiments, has been developed. Being compact and built with selected radio-pure materials, it is appropriate for low-background experiments. 
UV light generated by a pulsed lamp produces photoelectrons on a metallic cathode, which are subsequently accelerated to produce X-rays. 
The X-rays are emitted in bunches with a time resolution better than 1 ns
and the system provides a fast trigger for event tagging.
We present the idea of the new X-ray generator as well as
experimental results showing the proof of concept of the device. Data are in
good agreement with simulation results and future optimisations of the system
are also discussed.
\end{abstract}


\end{frontmatter}


%
%

\section{Introduction}
\label{intro}
An X-ray generator is an energy converter which receives electrical energy
and turns it into X-radiation. It is a relatively simple electrical device containing
two main elements: a cathode and an anode. In the standard device electrons
are produced on the cathode by thermionic effect and are then accelerated by a difference of voltage between two electrodes.
Electrons interact with the individual atoms of the anode material and produce either the characteristic fluorescence of the material
or Bremsstrahlung X-ray photons.

In most commercial X-rays generators electrons are produced in continuous mode by the thermionic effect
and the heat produced in this process is an undesirable by-product.
In the concept presented here electrons are generated by the irradation of a metal photocathode by pulsed extreme ultraviolet (EUV) photons, generated by an EUV lamp operated in pulsed mode. As its power consumption is very low ($\sim$ 1 mW), no heating up process is expected
and no special device to cool down the X-ray generator is needed.

The lamp also provides a signal to label with a $\sim$1 ns precision timestamp
the bunch of EUV photons and therefore the event tagging of the X-rays emission of X-rays.
This feature is of interest for any experiment requiring long-term data acquisition,
as the device can be continuously used for the detector calibration without interrupting
the data-taking.

In addition the dose rate should be well below 1 \textmu Sv/h at a distance of 10 cm,
and given that the high voltage is less than 30 kV,
this device should be exempt from special authorization \cite{RP:134},
simplifying its use in any installation.
Moreover, due to its compactness (only few cm length to produce X-rays of less than 8 keV)
and rather simple set-up, it is transportable to make on site detector calibrations and it can be 
easily integrated inside a detector shielding. 

Another favourable point is that it can be constructed with selected materials
to obtain sufficient radio-purity to satisfy the requirements in low-background
experiments. An example is the detector calibration in rare event searches,
like double beta decay~\cite{Agostini:2017jim}, dark matter particles~\cite{Baudis:2018bvr} or axions~\cite{Irastorza:2018dyq},
for which any radioactivity can produce background signals which can mimic the expected signal.
In general, all detectors installed in underground laboratories may profit from this device.

\label{Desription of the device}
\subsection{Description of the proposed new X-ray generator}

\begin{figure}[h!]
\centering
\includegraphics[width=0.90\textwidth]{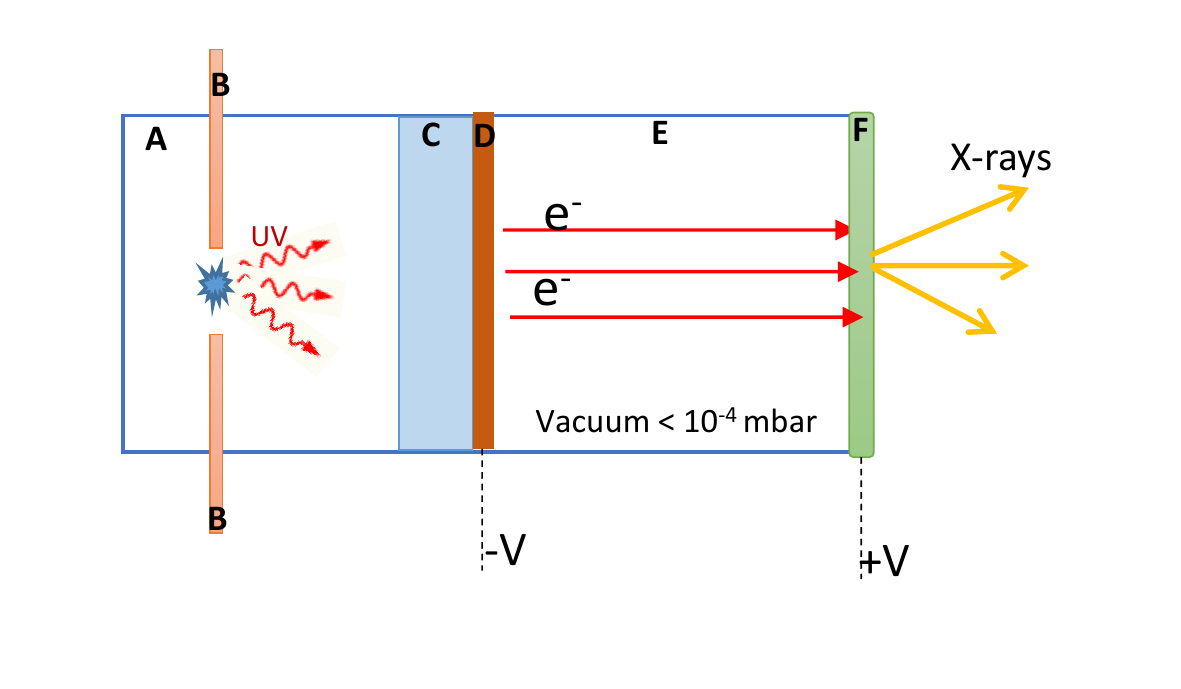}
\caption{Sectional view of the new X-ray generator: left to right is the UV lamp (A), its electrodes (B), the UV window (C) carrying the photocathode (D) and the target (F) where electrons are converted into X-rays. Electrons are accelerated in the gap (E) defined between the cathode (D) and the anode (F), in a vacuum better than 10$^{-4}$\,mbar.}
\label{fig:concept}      
\end{figure}

A sectional view of the new X-ray generator device is shown in Figure \ref{fig:concept}. The different components are: a UV gas discharge lamp, the UV window, the
photocathode and the X-ray converter. The process is as follows: the photons produced in the lamp are converted into electrons in the photocathode. These
electrons are then accelerated to hit the anode, where the X-rays are produced.

In this paper the proof of principle of the device is described. The full R\&D of this project also includes the manufacturing of the UV lamp,
the optimization for improving the temporal resolution, the light yield and the integration in a small set-up.
The compartment of the UV lamp is filled with a gas at a pressure lower than 1 bar. Different gases can be chosen, such as argon, xenon, deuterium
or hydrogen, to produce light at different wave-lengths (from 110 to 200 nm) and
with different timing performance (from 1 ns to 1.5 \textmu s). Electrodes dimmensions can be optimized to get a temporal resolution better than 1 ns.

The photocathode is deposited on a UV-window whose material should be chosen to have the best possible transmission for the UV light generated in the selected gas. The intensity of the UV photons and the repetition rate of these discharges are regulated by a simple RC circuit. The photoelectrons are then accelerated under the electric field created in the vacuum gap between the cathode and the anode, and are absorbed in the thin metallic anode, thus producing X-rays. This is a common process for other commercial generators. However having very smooth parallel metal plates allow to reduce the space between the two electrodes while applying the necessary high voltage. 
Moreover, to arrive to the desired electron energies, the voltage is applied in both electrodes reducing the needed dynamic range of each HV unit, that can be limited to \mbox{$<$ 30 kV}. These characteristics lead to a compact device design of few cm length and simplified device operation.

The construction materials of the X-ray generator target must be selected by their robustness and good compatibility with vacuum ($10^{-4}$ mbar) and by their high radio-purity.
For several rare event searches an acceptable emission value is 100 mBq/Kg~\cite{Alvarez:2012as,Abgrall:2016cct,Leonard:2017okt}.
Two good candidates fulfilling these requirements are CuC$_2$ copper \cite{Edelweiss} and 316Ti-NIRONIT stainless steel \cite{NEXT}. For the anode, which is thin (less than 10 \textmu m),
the radio-purity is less demanding but keeping a good vacuum mechanically will
be more difficult. For this reason the decoupling of the window and electron-to-photon
 converter is envisaged. The insulating part can be made of ceramics
such as boron nitride which satisfies both criteria while for the UV crystal a
possible candidate full-filling both requirements is MgF$_2$.

The new concept has been tested in the laboratory with results in line with expectations.
The experimental set-up is described in section \ref{sec:ExperimentalSetup},
while the results for three different targets are discussed in section \ref{sec:ExperimentalResults}.
In section \ref{sec:Simulation}, simulations made to optimize the target thickness in terms of X-ray production rate are presented. They have also helped to corroborate the experimental observations. 

\section{Experimental setup}
\label{sec:ExperimentalSetup}

\begin{figure}[h!]
\centering
\includegraphics[width=0.90\textwidth]{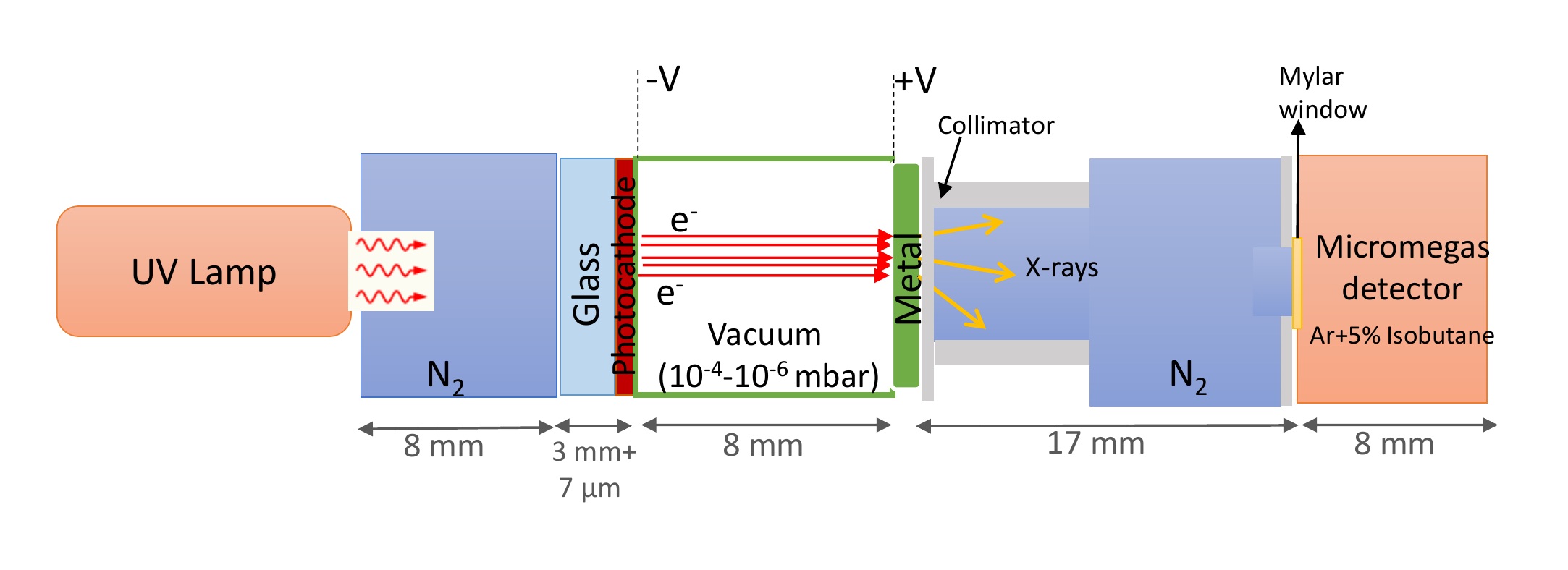}
\caption{Sketch of the experimental set-up, discribed in detail in the text.}
\label{fig:expSetup}      
\end{figure}

\begin{figure*}[t]
\centering
\includegraphics[width=0.85\textwidth]{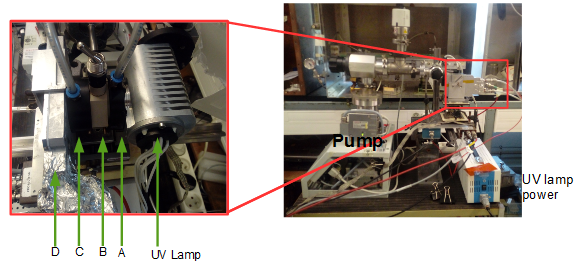}
\caption{Picture of the experimental set-up installed at CEA. On the right general view and on the left a detail outlook of the different elements that compose what is indeed the generator in this particular set-up: the UV lamp, the (A) volume 17\,mm thick filled with N$_{2}$, the vacuum region (B) where the photoelectrons are created (at the entrance), then accelerated (8\,mm length) and finally on the other side of the volume is the metallic layer to produce the X-rays. (C) is another N$_{2}$ volume to facilitate the transmission of the X-rays up to (D), that is the Micromegas detector.}
\label{fig:expSetupLab}      
\end{figure*}

The experimental set-up presented here as the proof of concept of the new
X-ray generator differs slightly from the general schema as it corresponded to initial tests.
A sketch of the set-up is shown in Figure \ref{fig:expSetup} and a picture in Figure \ref{fig:expSetupLab}.
The UV photons emitted from the lamp Hamamatsu L2D2 (type L10904)
cross a 8 mm long nitrogen volume arriving to a 7 \textmu m thick aluminium photocathode
deposited on 3 mm thick sapphire crystal.
Electrons are then accelerated in a 8 mm long vacuum region before arriving to the metallic anode, where they are
absorbed to produce radiation as characteristic Bremsstralung and X-rays.
The anode is attached to a collimator with a 6 mm diameter opening.
The generated X-rays are then detected in a Micromegas detector~\cite{MMs,MMs2}
after crossing another nitrogen 17 mm long volume. 

Different metallic windows have been used to produce X-rays, based on the desired spectrum and availability. 
They are summarized in Table \ref{tab:windows}. The vacuum reached varies from 10$^{-4}$ to 10$^{-6}$ mbar.
The applied voltage difference between the electrodes ranges from 2 kV up to 13 kV,
resulting in electron energies between 2 and 13 keV.

\begin{table*}[b!]
\centering
\caption{The three metallic targets tested in our experimental setup: thickness,
vacuum level reached in the electrode gap and generated characteristic x-rays, extracted from~\cite{X:Booklet}.}
\begin{tabular}{c|cc|ccc}
\hline
Material & Thickness & Vacuum & & Energy K$_\alpha$ & Energy K$_\beta$\\
         & (\textmu m) & (mbar) & & (keV) & (keV)\\
\hline
Aluminium       & 10 & $0.55 \times 10^{-5}$ & Al & 1.497 & \\
\hline
Stainless steel & 50 & $0.16 \times 10^{-5}$ & Cr & 5.410 & 5.947\\
                &    &                       & Mn & 5.890 & 6.490\\
                &    &                       & Fe & 6.400 & 7.058\\
                &    &                       & Ni & 7.472 & 8.265\\
\hline
Copper + Kapton          &  2 + 12.5 & $3.0  \times 10^{-5}$ & Cu & 8.041 & 8.905\\
\hline
\hline
\end{tabular}
\label{tab:windows}
\end{table*}

The Micromegas used was a microbulk Micromegas \cite{microbulk,Iguaz:2012ur} operating in Ar-5\%C$_4$H$_{10}$ in circulation mode. The detector is placed in an aluminium gas chamber with a hole of 3 mm diameter closed by an aluminized Mylar foil 4 $\mu$m thick to allow the X-rays to enter the gas volume. The chamber, 8 mm long, is grounded and the mylar window operates as the drift electrode while positive voltages are applied to the mesh and anode of the Micromegas. All data were acquired at the same operational point with a mesh voltage of 150\,V and drift voltage of 520\,V. The signals produced in the Micromegasgo through a preamplifier ORTEC 142-B and an ORTEC 572A amplifier, before being digitized by a Multi-Channel Analyzer (MCA) from Amptek and further analysed with a C++ routine based on ROOT \cite{ROOT}. The shaping time of the amplifier was set to a value of 12 $\mu$s.

\begin{figure}[t]
\centering
\includegraphics[width=9cm]{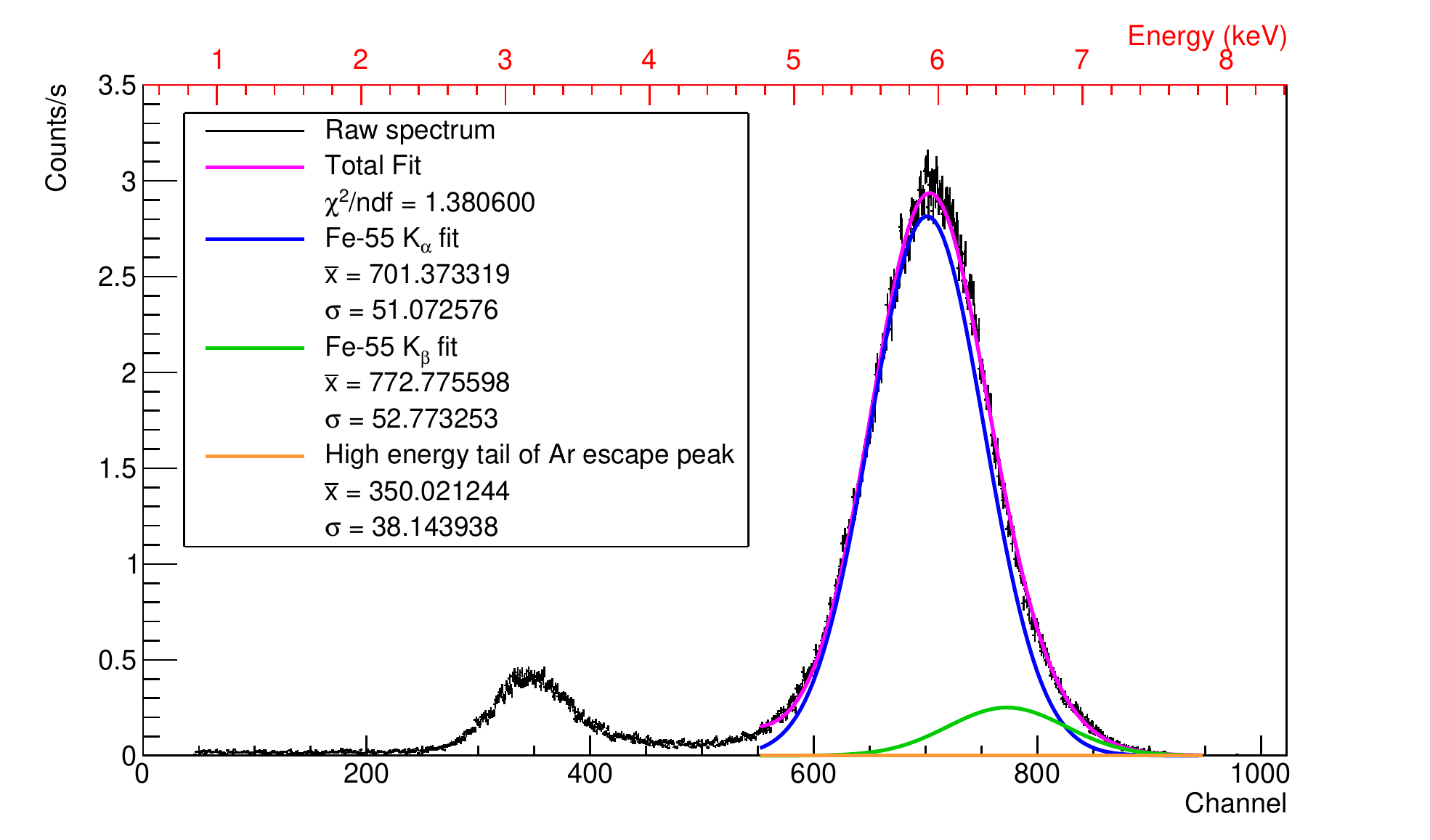}
\caption{$^{55}$Fe spectrum and its corresponding fit and energy calibration (red axis).}
\label{fig:Fe55}      
\end{figure}

The  detector has been calibrated by a $^{55}$Fe source (Figure \ref{fig:Fe55}). A fit to its k$_{\alpha}$ and k$_{\beta}$ lines (respectively at 5.9 and 6.4 keV) by two Gaussian functions is done assuming a linear background (first order polynomial). The escape peak of argon is also visible around channel 300. At this fields configuration the energy resolution obtained from the fits is of 18\% (FWHM) at 5.9 keV. 

\section{Experimental results}
\label{sec:ExperimentalResults}

Three electrons-to-photons targets have been tested in the proof of concept of the x-ray generator.
These targets play the role of the anode in the vacuum region where the electrons are accelerated.
They also act as upstream window closing this region and thus defining the achievable vacuum level.
Data have been acquired at different electron energies:
from 3.7 to 9.5 keV for the 10 \textmu m thick aluminium target,
from 6.3 to 9.5 keV for the 50 \textmu m thick stainless steel target
and from 10.5 to 13.5 keV for the 2 \textmu m thick copper target.

\begin{figure*}[hp!]

\begin{minipage}{.5\textwidth}
\centering
\includegraphics[width=1\textwidth]{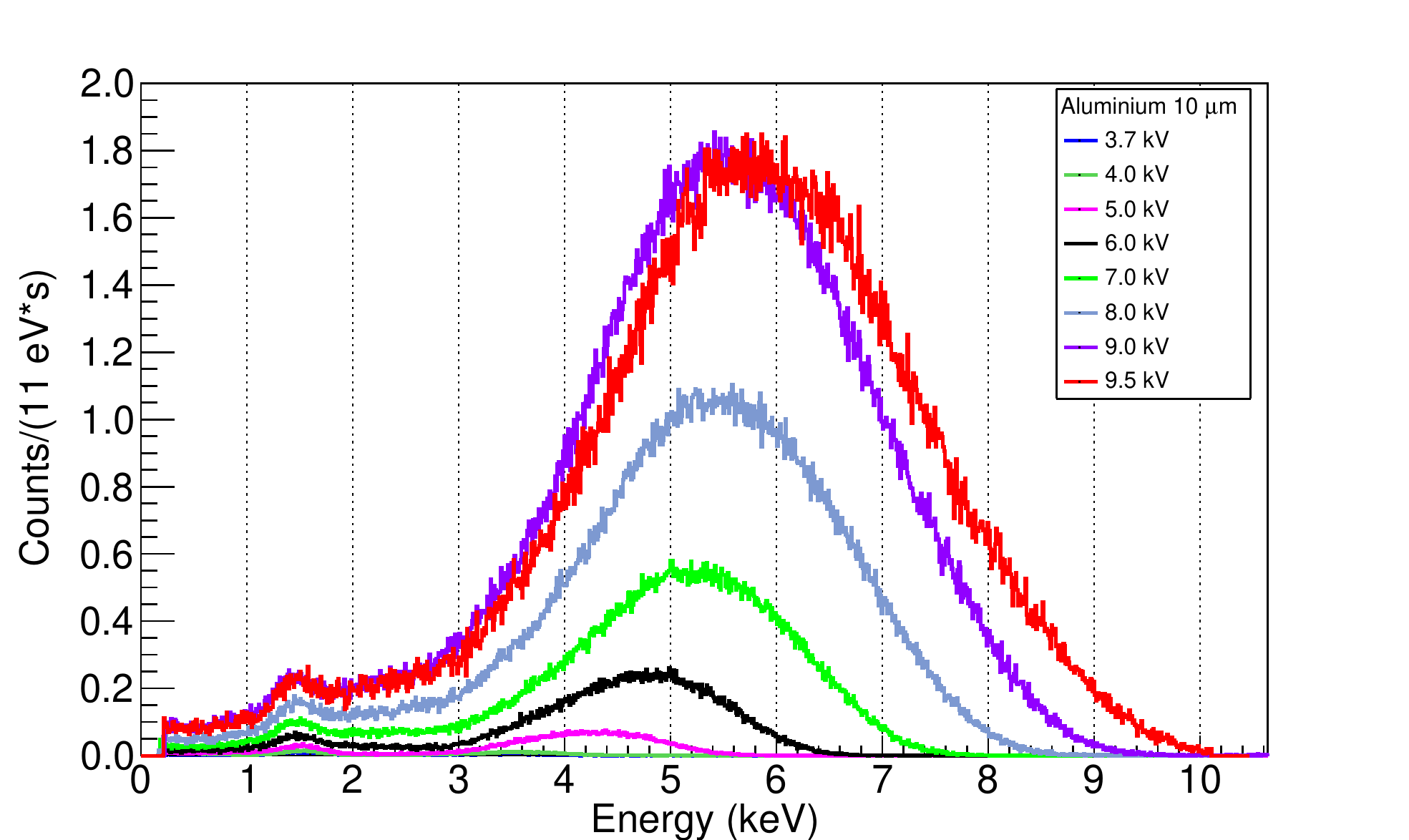}
\includegraphics[width=1\textwidth]{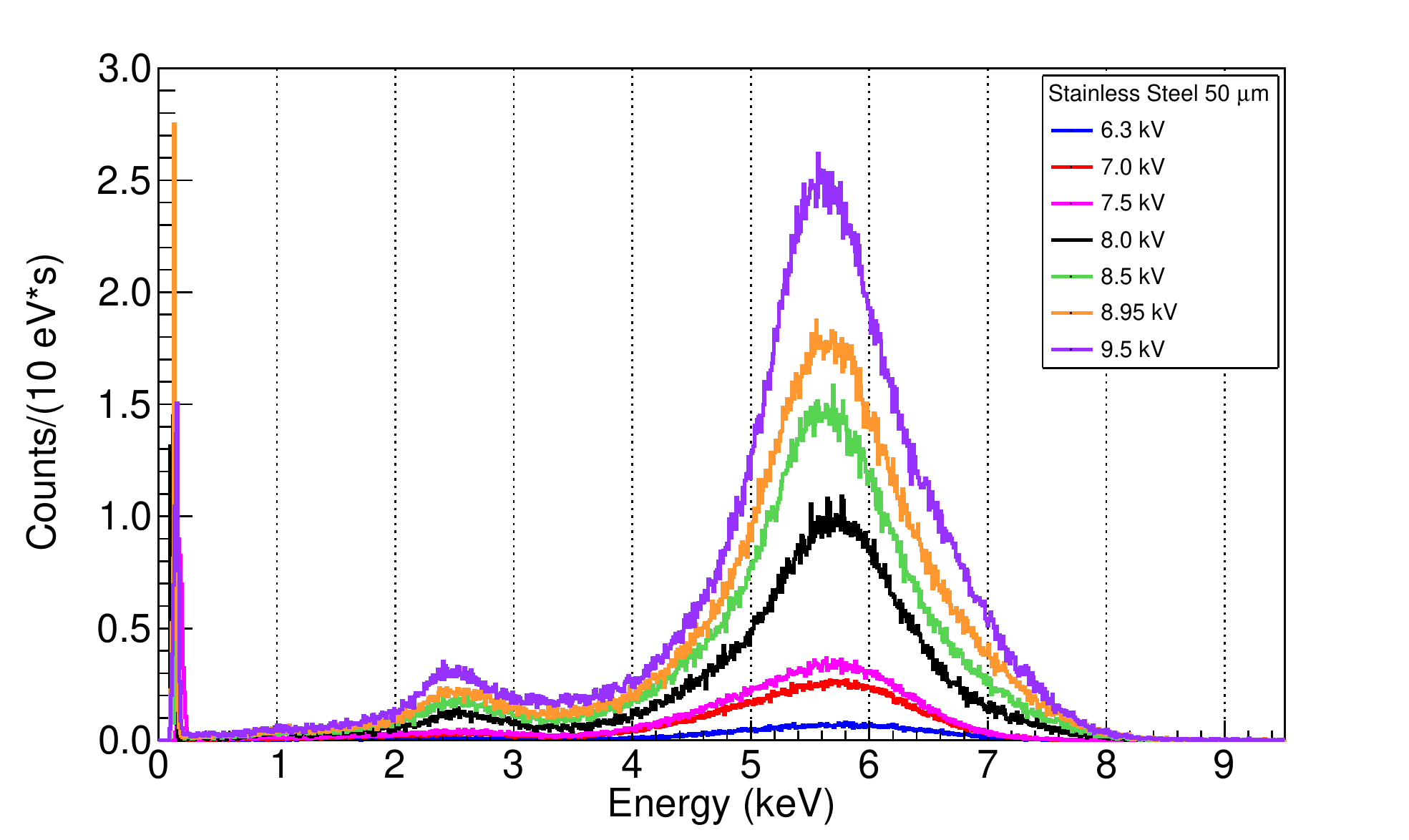}
\includegraphics[width=1\textwidth]{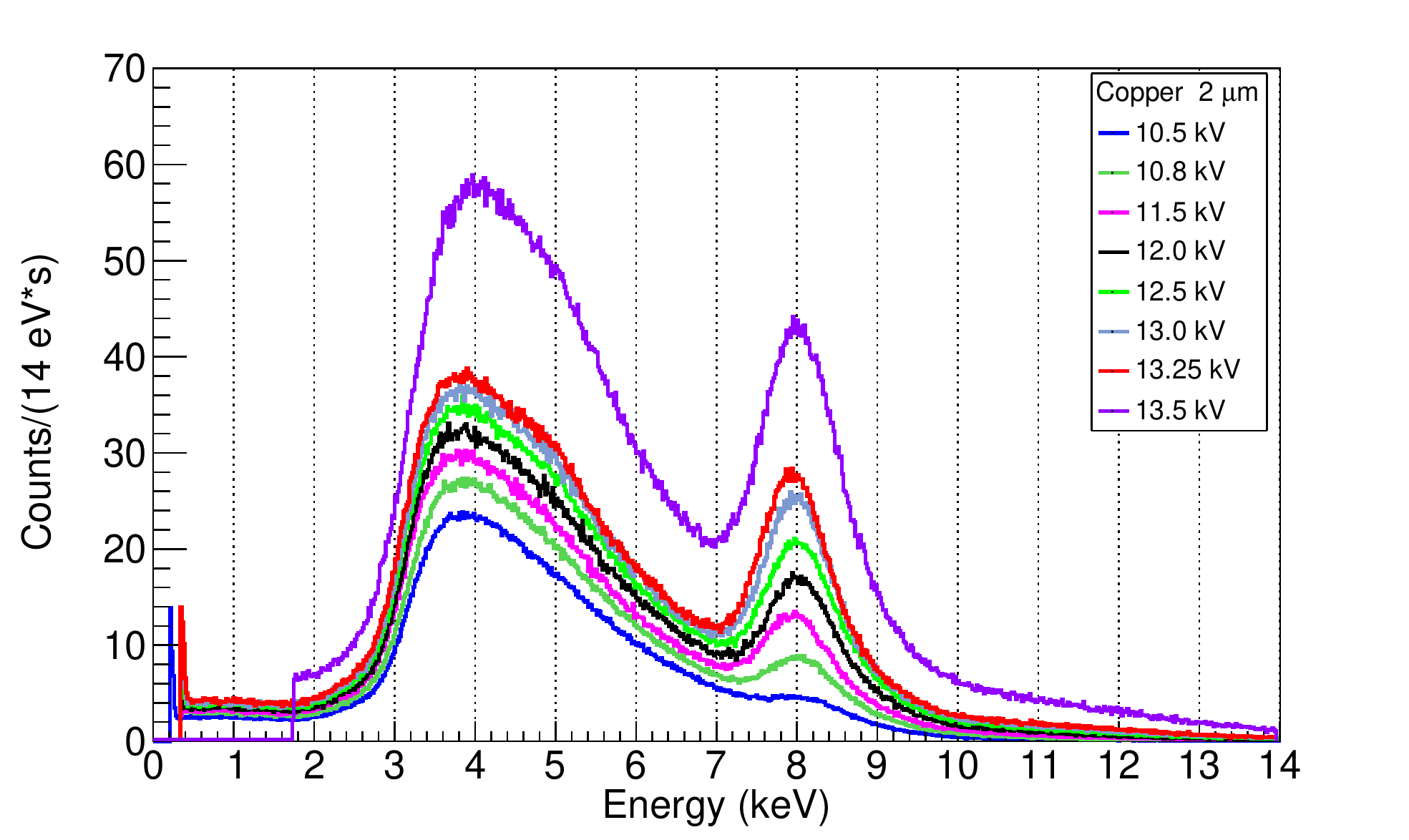}
\end{minipage}%
\begin{minipage}{0.5\textwidth}
\centering
\includegraphics[width=1\textwidth]{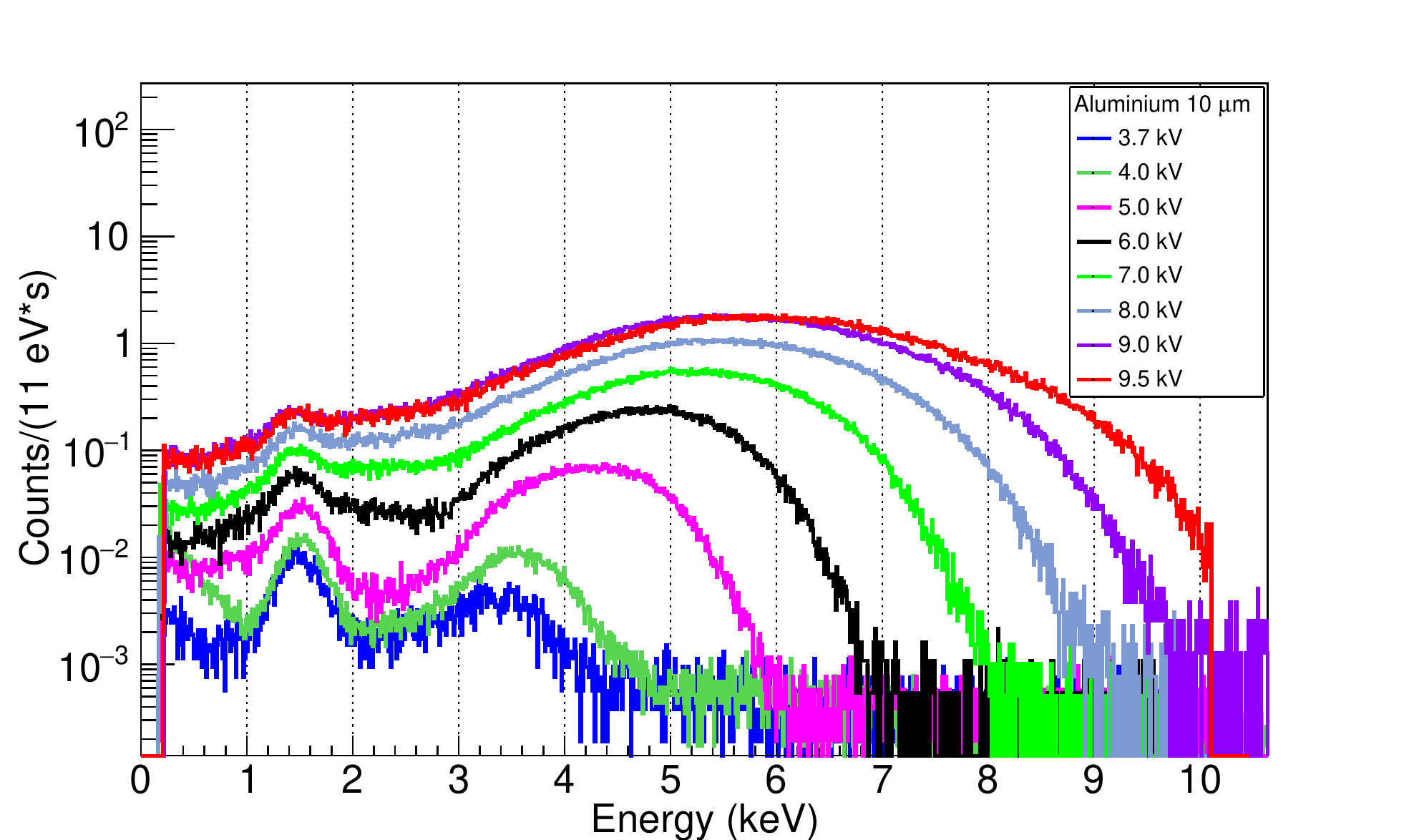}
\includegraphics[width=1\textwidth]{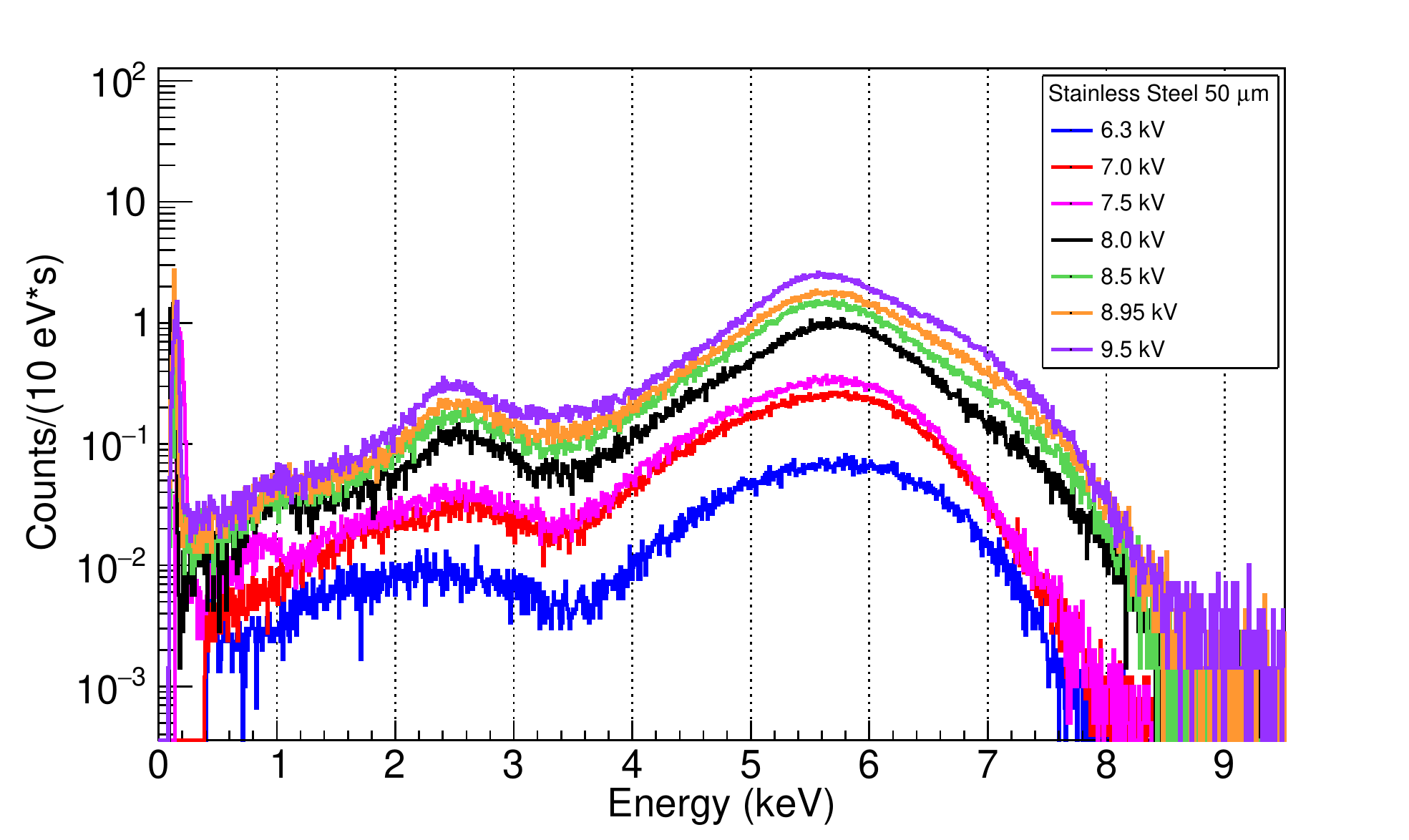}
\includegraphics[width=1\textwidth]{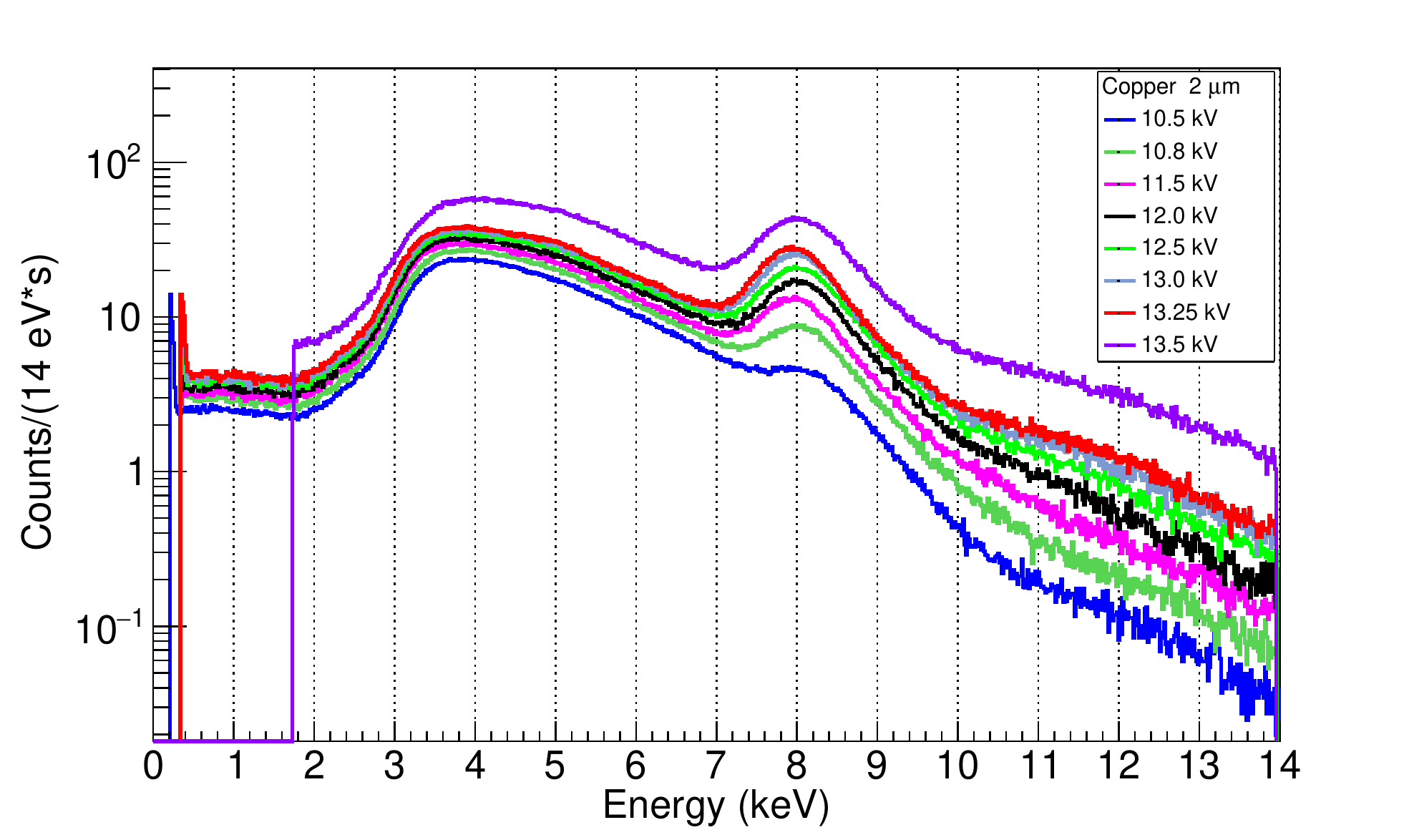}
 \end{minipage}
\caption{Energy spectra (in linear scale on the left and in logarithmic scale on the right) measured by the Micromegas detector of X-rays produced by the three tested targets:
10 \textmu m thick aluminium (top), 50 \textmu m thick stainless steel (middle)
and 2 \textmu m thick copper (bottom). Each spectrum has been normalized to the acquisition (exposure) time.}
\label{fig:ExpRes}
\end{figure*}

Each metal produces a different X-ray spectrum that depends on the window thickness
and on the initial electron energies that impinge on the window,
as shown in Figure \ref{fig:ExpRes}.

The main characteristic X-rays lines for the materials tested are listed in Table \ref{tab:windows},
where the Siegbahn notation has been adopted.
In almost all cases, only the k$_\alpha$ lines are visible:
the 1.5 keV line for aluminium,
several non-separable lines between 5.4 and 7.5 keV for stainless steel
and the 8.0 keV line for copper.
These lines emerges from a continuous Bremsstrahlung spectrum,
which ranges down to the self-absorption of the material (for instance, 2-3 keV for the aluminium)
up to the initial electron energy.
In the stainless steel and copper targets, an argon escape peak is also present,
produced by the non-absorption in the Micromegas detector of argon gas de-excitement photons,
with an energy of 2.9 keV.

\begin{figure}[t]
\centering
\includegraphics[width=0.70\textwidth]{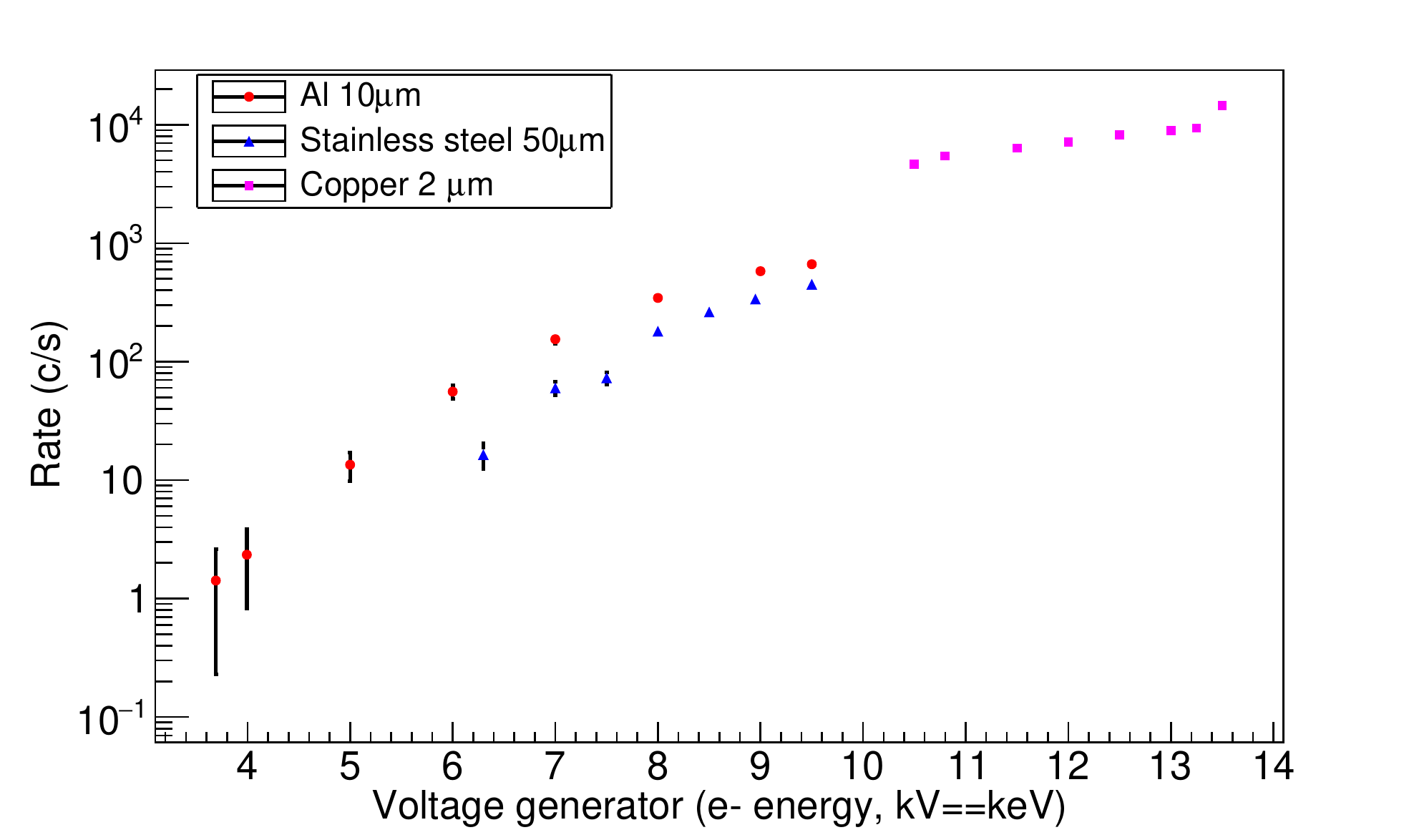}
\caption{Measured total rate in the Micromegas detector for the three tested targets: 10 $\mu$m thick aluminum (red circles), 50 $\mu$m stainless steel (blue triangles) and 2 $\mu$m thick copper (magenta squares). The statistical error is given for each point.}
\label{fig:ratesAll}     
\end{figure}

The total measured X-rays detection rate, proportional to the production one,
has been calculated for each electron energy
integrating each energy spectrum for an energy threshold of 0.5 keV.
Rates are shown in Figure \ref{fig:ratesAll} for the three metallic targets.
Detection rates up to kHz have been measured.
The metal thinkness can be optimized through MonteCarlo simulations
to minimize the  X-ray self-absorption in the target, keeping a high-vacuum.
First results of this work are shown in section \ref{sec:Simulation}.

\begin{figure}[!htb]
   \centering
   \begin{subfigure}{.55\textwidth}
   \centering
\hspace{-30pt}  \includegraphics*[width=1.1\columnwidth]{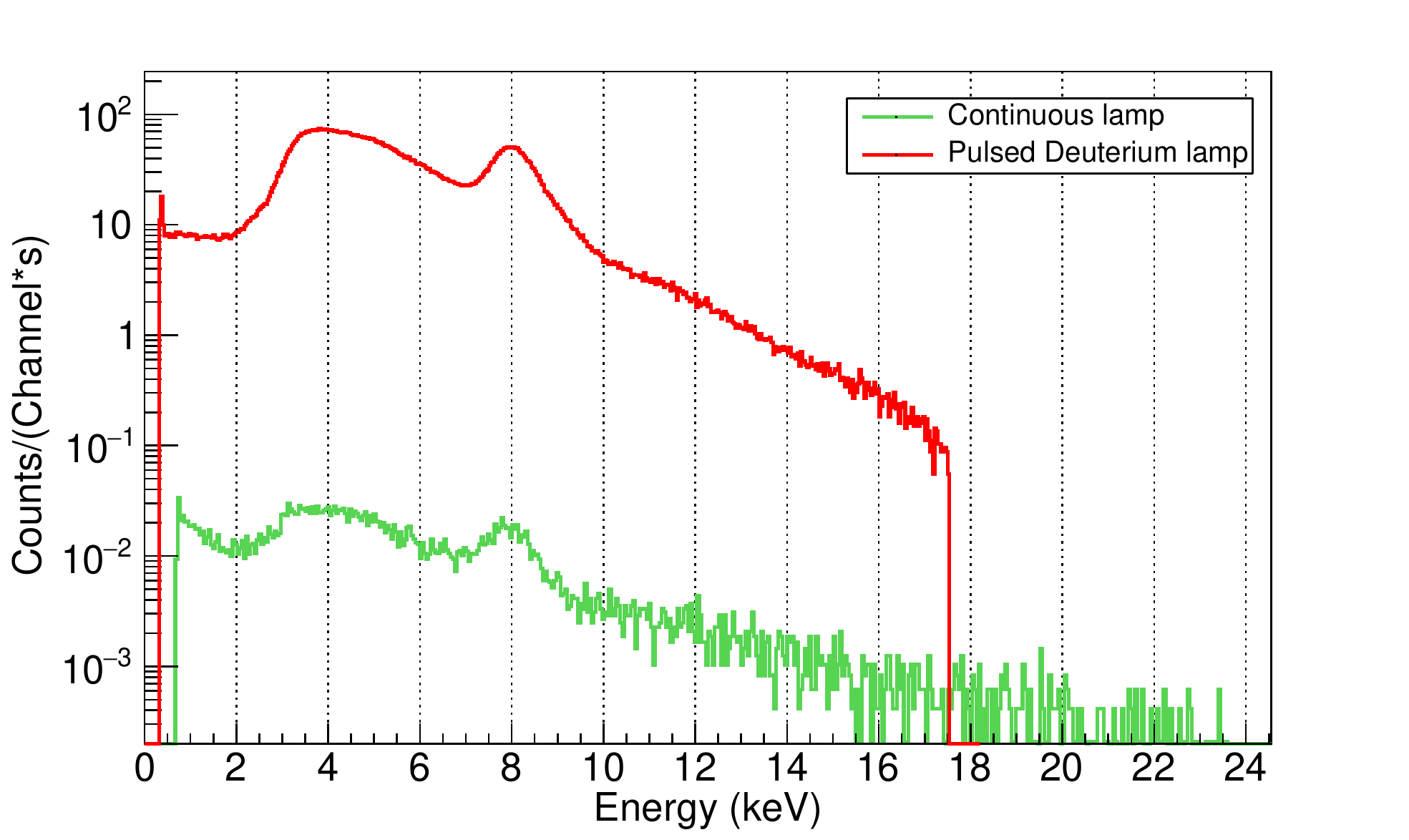}
   \caption{}
   \label{fig:Cupulsed}
   \end{subfigure}%
    \begin{subfigure}{.45\textwidth}
   \centering
\includegraphics*[width=1.0\columnwidth]{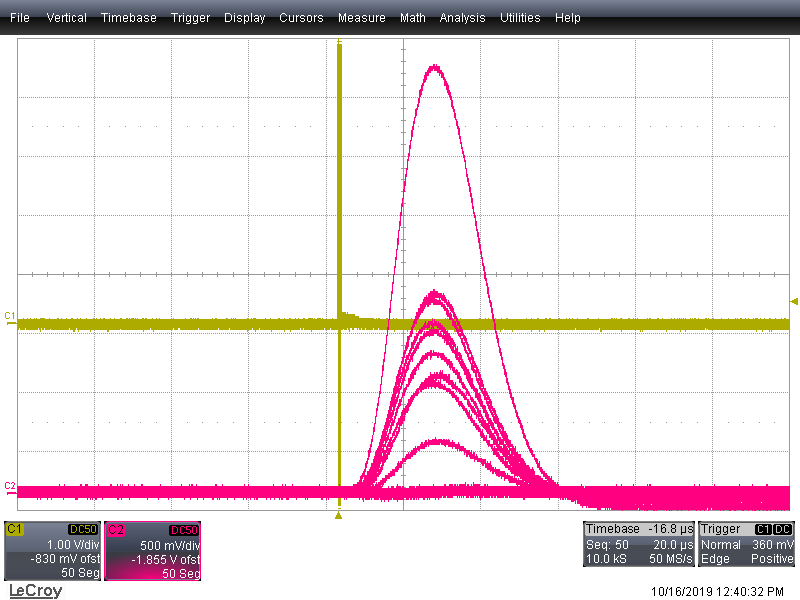}
   \caption{}
   \label{fig:cupulse}
   \end{subfigure}%
\caption{(a) Energy spectra from the copper target detected in the Micromegas. In red are the data from the continuous lamp and in green those from the pulsed deuterium lamp. (b) Event pulses registered in with an oscilloscope (magenta) with trigger given by the pulsed lamp (yellow signal).}
\label{fig:Cu_pulsed_lamp}     
\end{figure}

Additional data were acquired for the copper target replacing the UV continuous lamp
by a hydrogen pulsed lamp, remaining the rest of the set-up exactly the same.
The lamp was operated at a voltage between the electrodes of 2.5 kV.
A Cu spectrum is shown in Figure \ref{fig:Cu_pulsed_lamp} (left) for a given electron acceleration.
On the right side of the same figure, events registered with an oscilloscope (in magenta) in
coincidence (yellow line) with the trigger of the lamp are shown. In this case the
rate is much lower because the lamp repetition rate was 20 Hz with a measured
detection probability of 20\%. However, the tests proved the feasibility of the
final set-up with the integration of a pulsed lamp.

\begin{figure*}[hp!]
\begin{minipage}{0.3\textwidth}
\centering
\hspace*{-0.5cm}\includegraphics[width=1.3\textwidth, height=1.1\textwidth]{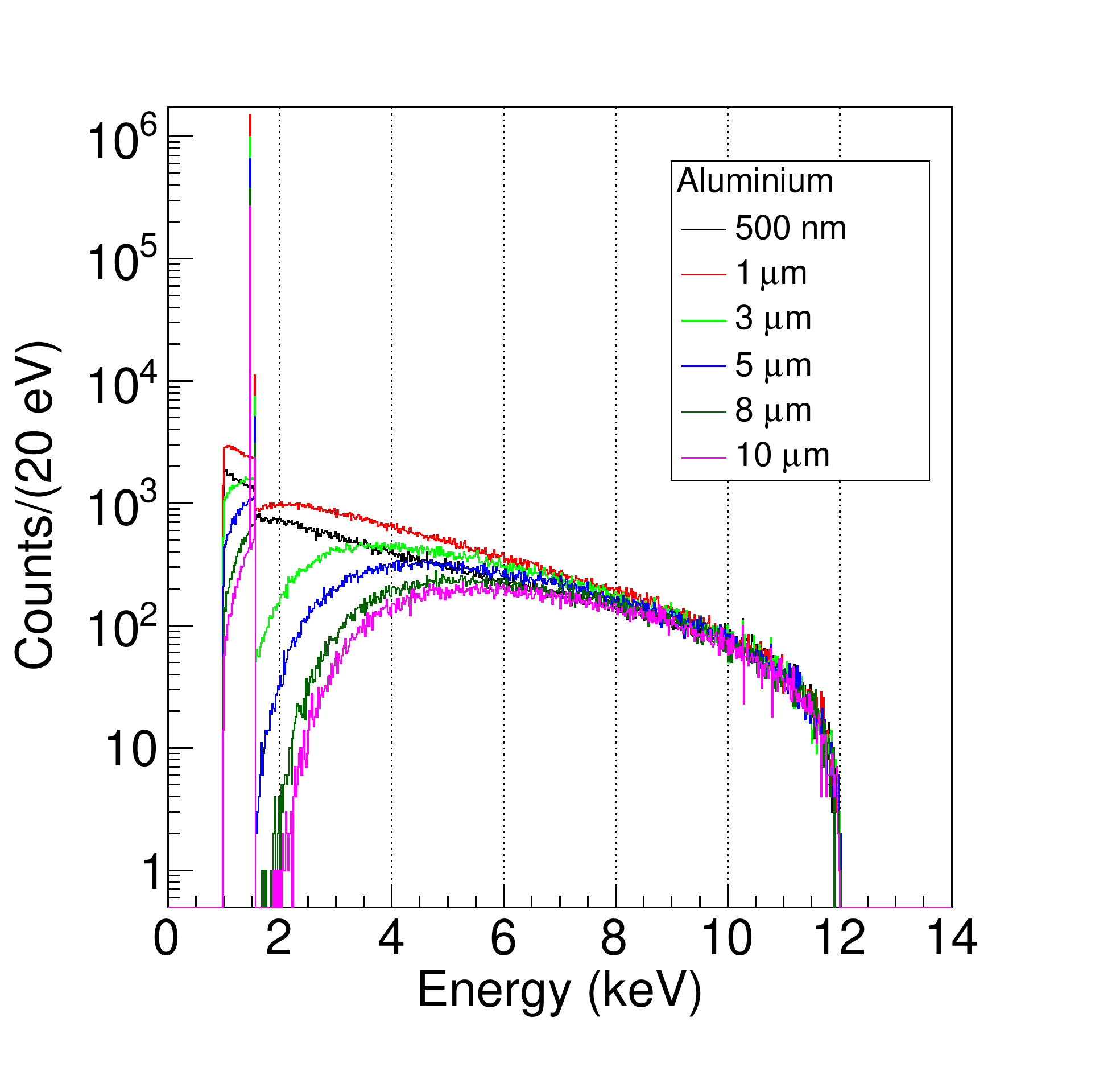}
\end{minipage}
\begin{minipage}{0.3\textwidth}
\includegraphics[width=1.3\textwidth, height=1.1\textwidth]{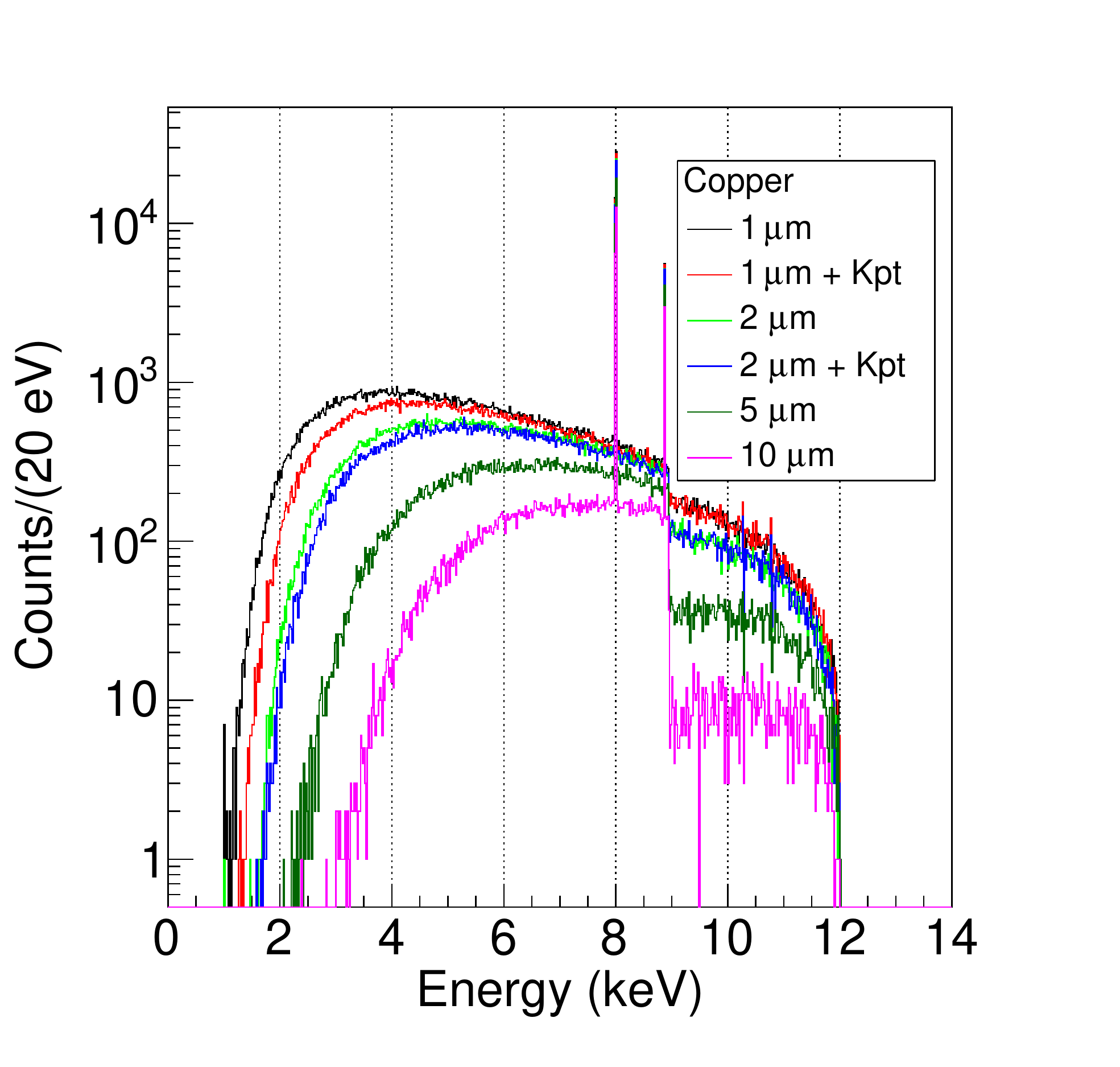}
\end{minipage}
\begin{minipage}{0.3\textwidth}
\hspace*{0.5cm}\includegraphics[width=1.3\textwidth, height=1.1\textwidth]{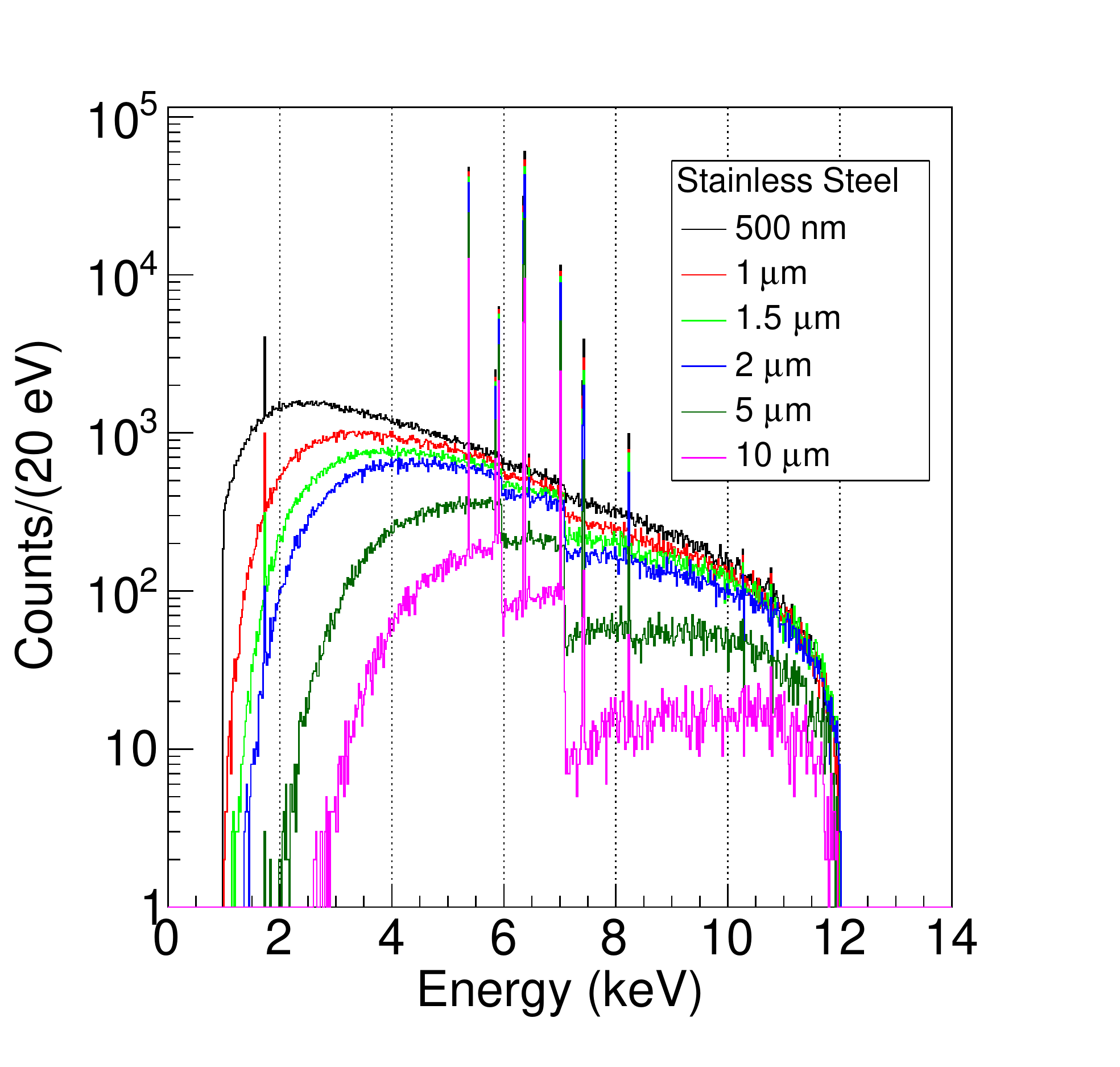}
\end{minipage}
\caption{GEANT4 simulated x-ray energy spectra produced by 12 keV electrons in different thicknesses of 
aluminium (top), stainless steel (middle) and copper (bottom). All spectra are normalized to the same number of incident electrons.}
\label{fig:Sim12keV}
\end{figure*}

\section{Simulation program and geometry optimisation}
\label{sec:Simulation}

The X-ray spectrum produced in the electron-to-photon conversion process, the X-ray yield and the energy deposited by these photons in a microbulk were simulated by means of the GEANT4 toolkit \cite{geant4}. Two different geometries were implemented for these purposes: a minimal one, including only the electron-to-photon coverter and a region in front of it, and a complete one, comprehensive of the whole set-up used experimentally. 
In the minimal one, a monochromatic pencil-like electron beam was defined in the vacuum region just before the converter and the X-ray spectrum was scored at its exit surface. This simulation was performed for the same three metallic converters tested experimentally, but for more thicknesses. For the inox material composition, stainless steel 304-L \cite{MCNP} was considered. The incident electron energy was 12 keV. The results are reported in Fig. \ref{fig:Sim12keV}, where the threshold of 990 eV corresponds to the default cut-off energy for X-rays in GEANT4. As  expected,  the main characteristics X-ray lines reported in Table \ref{tab:windows} (plus the 1.7 keV k$\alpha$ line from Si in Inox) emerge from a Bremmstrahlung extending up to 12 keV.  By tuning the binning, more  lines could be resolved. 
From Fig.  \ref{fig:Sim12keV} it is also evident that, the thicker the foil, the higher the self-absorption (therefore the lower the transmission) of the photons produced.

\begin{figure}[hp!]
\centering
\includegraphics[width=0.70\textwidth]{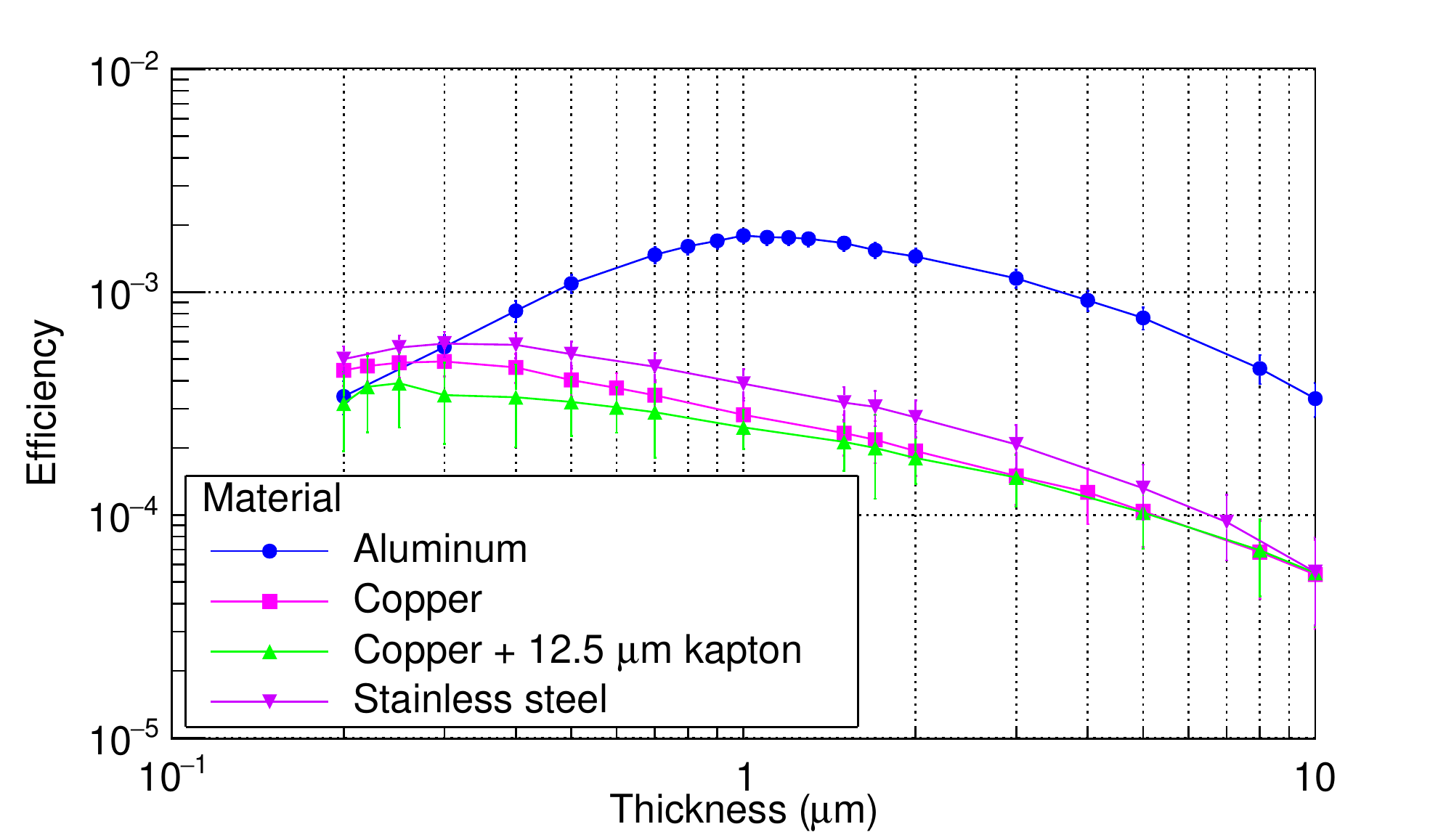}
\caption{GEANT4 simulated X-ray production yield for 12 keV electrons against different thickness of aluminum (blue circles), copper (magenta squares), copper followed by 12.5 $\mu$m kapton (green triangles) and stainless steel (violet reverse triangles).}
\label{fig:SimEff}     
\end{figure}

Fig. \ref{fig:SimEff} shows the photon production efficiency calculated from these simulations for the three electron-to-photon converters, quantified as the number of X-rays per incident 12 keV electrons emerging from the conversion layer. The optimal thickness is of the order of the one absorption length of the electrons in the material:  if the foil is too thin not all electrons are stopped inside it, while if it is too thick self absorption takes over. From the copper points it is visible that the presence of the kapton foil reduces the number of X-rays emerging from the metal. The reduction is more important for thinner metallic layers, passing from 12\% for Cu foils of 1  $\mu$m thick to 6\% for Cu foils 2 microns thick. 

\begin{figure}[h]
\centering
\includegraphics[width=9cm]{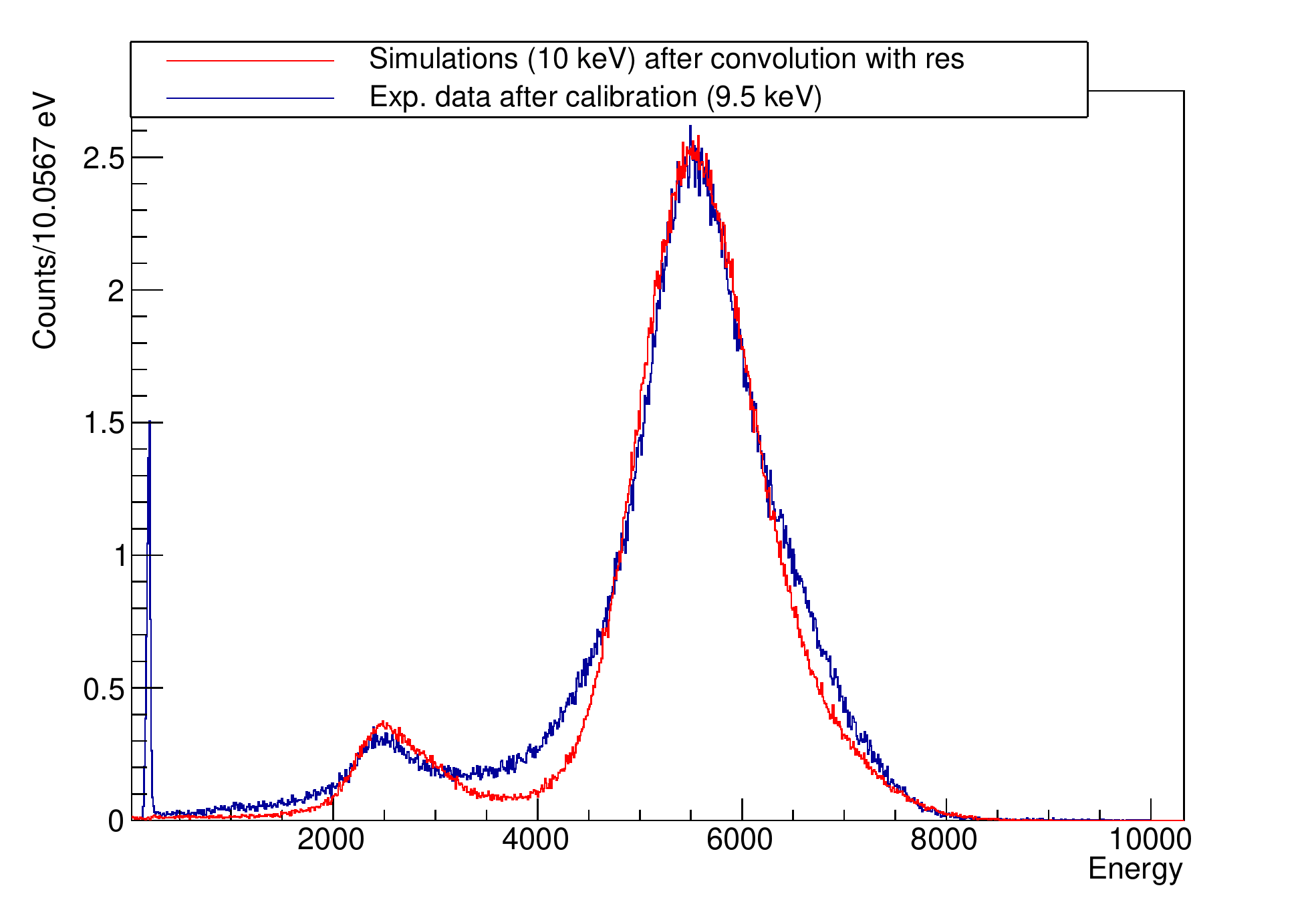}
 \caption{Comparison between experimental data and GEANT4 simulated energy deposited in the sensitive gas volume of the microbulk when a 10\,keV monochromatic electron beam impinges on a 50\,$\mu$m inox foil. The resolution of the detector has been accounted for.}
 \label{InoxExpSim}      
 \end{figure}

Thought Fig \ref{fig:SimEff} agrees with the expectations, it can not be directly compared with the experimental energy spectra because large fraction of the set-up was unaccounted for in the simulations and the spectra were not obtained at the same location.  On the other hand, if the electron-to-photon converter is thick, performing a single simulation with the whole geometry would be too time consumming. Therefore to be able to compare the experimental data to the simulations the previously introduced geometry was  used to produce the X-ray spectrum obtained when a monochromatic electron beam impinges on a foil thick enough that no electron is expected to emerge.
The spectrum is then input in the whole geometry, in front of an additional layer of the same electron-to-photon converter.  In the latter geometry the scoring region is the Micromegas gas volume and the tallied quantity the deposited energy.  In the specific case, in the minimal geometry the electron energy was 10 keV and  the stainless steel layer measured 10  $\mu$m. The inox thickness was then set to  40  $\mu$m in the complete geometry, adding up to a total of 50 $\mu$m.\\
The result, after convolution with the energy resolution of the detector (18\,\% FWHM, Figure \ref{fig:Fe55}), is shown in Figure \ref{InoxExpSim} in comparison with the experimental results obtained for a similar field. The peak around 6\,keV accounts for all the characteristic X-rays listed in Table \ref{tab:windows}. Around 2.5\,keV, the Argon escape peak of the unresolved lines is visible. The spectra, normalised to the same peak height, are quite in agreement. The small difference between the two spectra is seen in between the two peaks and is due to imperfections in the microbulk structure which have not been taken into account in the simulations.

\section{Conclusions and perspectives}

A novel type of X-ray generator has been built. A simplified design of the device was tested in the laboratory, validating the proof of concept.    
GEANT4 simulations showed a good agreement with experimental results, opening the way to geometry and material optimisation. 


The simplicity of the proposed instrument makes it appropriate for various applications. We highlight the following:
\begin{itemize}
    \item The device is operating at moderate voltages ($\leq$ 20 kV ) and very low consumption (1 mWatt) thanks to the pulsed mode and the use of UV light for the electron extraction instead of thermionic emission. Thus there is not need of special authorisation or safety equipment, while the whole apparatus could operate with batteries or solar cells. Moreover the apparatus is very compact and light. As a result it is a portable calibration system that can be easily installed close to the detector, overcoming the necessity of bringing the detector to an X-ray generator facility.
    \item Our generator can be permanently installed next to detectors requiring a high radio-purity as in the case of dark matter or neutrino experiments, or applications as the radiometric dating of materials studying traces of radioactive impurities.
    \item The device can be used to calibrate detectors during data collection without interruption of the experiment thanks to the tagging of the events. This is very important for low-rate, rare-event experiments. However even in case of high-rate experiments tagging is still possible thanks to the excellent timing of the device (given the UV lamp resolution is .35 ns we expect an overall jitter $\leq$ 1ns for the X-ray buckets). 
\end{itemize}

\section{Acknowledgments}
F. J. Iguaz acknowledges the support from the Enhanced Eurotalents program (PCOFOUND-GA-2013-600382).

\color{black}
\newpage

\section*{References}
\bibliography{mybibfile}

\begin{thebibliography}{10}
\expandafter\ifx\csname url\endcsname\relax
  \def\url#1{\texttt{#1}}\fi
\expandafter\ifx\csname urlprefix\endcsname\relax\def\urlprefix{URL }\fi
\expandafter\ifx\csname href\endcsname\relax
  \def\href#1#2{#2} \def\path#1{#1}\fi

\bibitem{RP:134}
M.~Y. Gerchikov, Z.~K. Hillis, E.~I.~M. Meijne, W.~Oatway, S.~Mobbs, A.~van
  Weers, Evaluation of the application of the concepts of exemption and
  clearance for practices according to title {III} of {Council} {Directiv}e
  96/29/{EURATOM} of 13 {May} 1996 in {EU} {Member} {States},
  \url{https://ec.europa.eu/energy/sites/ener/files/documents/134_appendice.pdf},
  {Volume 2}, {Appendices} (2003).

\bibitem{Agostini:2017jim}
M.~Agostini, G.~Benato, J.~Detwiler, {Discovery probability of next-generation
  neutrinoless double-beta decay experiments}, Phys. Rev. D96~(5) (2017)
  053001.
\newblock \href {http://arxiv.org/abs/1705.02996} {\path{arXiv:1705.02996}},
  \href {http://dx.doi.org/10.1103/PhysRevD.96.053001}
  {\path{doi:10.1103/PhysRevD.96.053001}}.

\bibitem{Baudis:2018bvr}
L.~Baudis, {The Search for Dark Matter }, European Review. 26 (2018) 70--81.
\newblock \href {http://dx.doi.org/10.1017/S1062798717000783}
  {\path{doi:10.1017/S1062798717000783}}.

\bibitem{Irastorza:2018dyq}
I.~G. Irastorza, J.~Redondo, {New experimental approaches in the search for
  axion-like particles}, Prog. Part. Nucl. Phys. 102 (2018) 89--159.
\newblock \href {http://arxiv.org/abs/1801.08127} {\path{arXiv:1801.08127}},
  \href {http://dx.doi.org/10.1016/j.ppnp.2018.05.003}
  {\path{doi:10.1016/j.ppnp.2018.05.003}}.

\bibitem{Alvarez:2012as}
V.~Alvarez, et~al., {Radiopurity control in the NEXT-100 double beta decay
  experiment: procedures and initial measurements}, JINST 8 (2013) T01002.
\newblock \href {http://dx.doi.org/10.1088/1748-0221/8/01/T01002}
  {\path{doi:10.1088/1748-0221/8/01/T01002}}.

\bibitem{Abgrall:2016cct}
N.~Abgrall, et~al., {The Majorana Demonstrator radioassay program}, Nucl.
  Instrum. Meth. A828 (2016) 22--36.
\newblock \href {http://dx.doi.org/10.1016/j.nima.2016.04.070}
  {\path{doi:10.1016/j.nima.2016.04.070}}.

\bibitem{Leonard:2017okt}
D.~S. Leonard, et~al., {Trace radioactive impurities in final construction
  materials for EXO-200}, Nucl. Instrum. Meth. A871 (2017) 169--179.
\newblock \href {http://dx.doi.org/10.1016/j.nima.2017.04.049}
  {\path{doi:10.1016/j.nima.2017.04.049}}.

\bibitem{Edelweiss}
E.~Armengaud, et~al., Background studies for the edelweiss dark matter
  experiment, Astroparticle Physics 47 (2013) 1 -- 9.
\newblock \href
  {http://dx.doi.org/https://doi.org/10.1016/j.astropartphys.2013.05.004}
  {\path{doi:https://doi.org/10.1016/j.astropartphys.2013.05.004}}.

\bibitem{NEXT}
V.~{\'{A}}lvarez, et~al., Radiopurity control in the {NEXT}-100 double beta
  decay experiment: procedures and initial measurements, Journal of
  Instrumentation 8~(01) (2013) T01002--T01002.
\newblock \href {http://dx.doi.org/10.1088/1748-0221/8/01/t01002}
  {\path{doi:10.1088/1748-0221/8/01/t01002}}.

\bibitem{MMs}
Y.~Giomataris, P.~Rebourgeard, J.~P. Robert, G.~Charpak, {MICROMEGAS: A High
  granularity position sensitive gaseous detector for high particle flux
  environments}, Nucl. Instrum. Meth. A376 (1996) 29--35.
\newblock \href {http://dx.doi.org/10.1016/0168-9002(96)00175-1}
  {\path{doi:10.1016/0168-9002(96)00175-1}}.

\bibitem{MMs2}
Y.~Giomataris, {Development and prospects of the new gaseous detector
  Micromegas}, Nucl. Instrum. Meth. A419 (1998) 239--250.
\newblock \href {http://dx.doi.org/10.1016/S0168-9002(98)00865-1}
  {\path{doi:10.1016/S0168-9002(98)00865-1}}.

\bibitem{X:Booklet}
A.~Thompson, et~al., X-ray {Data} {Booklet},
  \url{https://xdb.lbl.gov/xdb-new.pdf} (September 2009).

\bibitem{microbulk}
S.~Andriamonje, et~al., {Development and performance of Microbulk Micromegas
  detectors}, JINST 5 (2010) P02001.
\newblock \href {http://dx.doi.org/10.1088/1748-0221/5/02/P02001}
  {\path{doi:10.1088/1748-0221/5/02/P02001}}.

\bibitem{Iguaz:2012ur}
F.~J. Iguaz, E.~Ferrer-Ribas, A.~Giganon, I.~Giomataris, {Characterization of
  microbulk detectors in argon- and neon-based mixtures}, JINST 7 (2012)
  P04007.
\newblock \href {http://dx.doi.org/10.1088/1748-0221/7/04/P04007}
  {\path{doi:10.1088/1748-0221/7/04/P04007}}.

\bibitem{ROOT}
R.~Brun, F.~Rademakers, Root - an object oriented data analysis framework,
  Nucl. Inst. and Meth. in Phys. Res. A 389 (1997) 81--86.

\bibitem{geant4}
S.~Agostinelli, \emph{et al.}, Nuclear and Instrument Methods A 506 (2003)
  250--303.

\bibitem{MCNP}
R.~McConn, C.~J. Gesh, R.~T. Pagh, R.~A. Rucker, R.~G. Williams, Compendium of
  material composition data for radiation transport modeling, revision 1,
  PNNL-15870 (2011).

\end{thebibliography}

\end{document}